\documentclass{article}
\usepackage{arxiv}
\usepackage[utf8]{inputenc} 
\usepackage[T1]{fontenc}    
\usepackage{hyperref}       
\usepackage{url}         
\usepackage{booktabs}    
\usepackage{amsfonts}      
\usepackage{nicefrac}      
\usepackage{microtype}     
\usepackage{lipsum}
\usepackage{float}
\usepackage{graphicx}
\usepackage{caption}
\usepackage{subcaption}
\usepackage{titlesec}
\usepackage{multirow}
\usepackage{amsmath}
\usepackage{soul,color}
\usepackage[section]{placeins}
\usepackage{verbatim}

\usepackage[nodisplayskipstretch]{setspace}

\title{Computational discovery of new 2D materials using deep learning generative models}
\author{
  Yuqi Song, Edirisuriya M. Dilanga Siriwardane, Yong Zhao\\
  Department of Computer Science and Engineering\\
  University of South Carolina\\
  Columbia, SC 29201 \\
    \And
 Jianjun Hu *\\
 Department of Computer Science and Engineering\\
  University of South Carolina\\
  Columbia, SC 29201 \\
  \texttt{jianjunh@cse.sc.edu} \\
}
\titlespacing {\section}
{0pt}{5.5ex plus 1ex minus .2ex}{3.3ex plus .2ex}
\begin{document}
\maketitle
\begin{abstract}

Two dimensional (2D) materials have emerged as promising functional materials with many applications such as semiconductors and photovoltaics because of their unique optoelectronic properties.
While several thousand 2D materials have been screened in existing materials databases, discovering new 2D materials remains to be challenging. Herein we propose a deep learning generative model for composition generation combined with random forest based 2D materials classifier to discover new hypothetical 2D materials. Furthermore, a template based element substitution structure prediction approach is developed to predict the crystal structures of a subset of the newly predicted hypothetical formulas, which allows us to confirm their structure stability using DFT calculations. So far, we have discovered 267,489 new potential 2D materials compositions and confirmed twelve 2D/layered materials by DFT formation energy calculation. Our results show that generative machine learning models provide an effective way to explore the vast chemical design space for new 2D materials discovery. 

\end{abstract}

\section{Introduction}

Two dimensional materials such as graphene and hexagonal boron nitride have the potential to create new electronics and technologies such as spintronics, catalysis, and membranes owing to their exotic vibrational, electronic, optical\cite{wang2019recent}, magnetic, and topological behaviors\cite{gibertini2019magnetic,akinwande2019graphene,ahn20202d,zhang2019high}. Using DFT based screening, Mounet et al. \cite{mounet2018two} have found 1,825 compounds with requisite geometric and bonding criteria that should make them relatively easy to exfoliate and so produce novel 2D materials with potentially interesting physical and electromagnetic properties. They discovered 56 ferromagnetic and antiferromagnetic systems, including half-metals and half-semiconductors. This greatly expands the list of predicted 2D materials and could fill the gaps in the characteristics and properties of the likes of graphene, phosphorene, and silicene. One can also expand the list of 2D materials by chemical substitutions, alternative site decorations, crystal structure prediction and so on\cite{paul2017computational}. Several screening approaches have been proposed to find 2D materials from known layered bulk materials \cite{choudhary2017high}. A simple criterion of comparing experimental lattice constants and lattice constants mainly obtained from Materials-Project (MP) density functional theory (DFT) calculation repository is used to find potential 2D materials \cite{choudhary2017high}: a relative difference between the two lattice constants for a specific material is greater than or equal to 5\% is used to identify good candidates for 2D materials. Haastrup et al. \cite{haastrup2018computational} developed the Computational 2D Materials Database (C2DB), which contains a variety of structural, thermodynamic, elastic, electronic, magnetic, and optical properties of around 1500 2D materials distributed over more than 30 different crystal structures. More recently, Zhou et al. \cite{zhou20192dmatpedia} developed 2DMatPedia, an open computational database of 6351 two-dimensional materials by screening all bulk materials in the database of Materials Project for layered structures by a topology-based algorithm and theoretically exfoliating them into monolayers. New 2D materials have also been generated by chemical substitution of elements in known 2D materials by others from the same group in the periodic table. These databases of experimental or hypothetical 2D materials have made it possible for discovering novel function materials \cite{fronzi2019impressive,yang2020high,lu2020coupling,kabiraj2020high,olsen2019discovering}.

Despite these efforts, the scale of experimental and hypothetical 2D materials is still limited. For example, computational generation of novel new materials have been proposed in the name of inverse materials design \cite{zunger2018inverse}, in which new materials are to be searched to achieve a given specific function, most of these methods involve a global optimization or search/sampling procedure to explore the search space\cite{chen2019smart}. However, most of such inverse design research is based on screening known materials. Suleyman Er et al. \cite{sorkun2020artificial}  proposed an elemental substitution based approach and applied it to known 2D materials structural prototypes to generate a large number of hypothetical 2D materials, and then filtered those materials based on several criteria. They deposited their predicted 2D materials in their V2DB database.

To expand the scope of 2D materials, we propose to design a generative deep learning method to discover novel 2D materials in uncharted composition space. Our approach is based on a high-accuracy composition based 2D materials classifier, which is used to screen millions of hypothetical materials compositions generated using our MatGAN, a generative adversarial network (GAN) based model  \cite{dan2020generative} that learns to generate chemically valid hypothetical materials. Based on 2.65 million generated samples, we have identified 267,489 hypothetical 2D materials.

Our contributions can be summarized as follows:

\begin{itemize}
  \item We propose a composition based 2D materials classifier model which achieves high prediction accuracy when trained with known 2D materials.
  \item We combine the 2D materials classifier and the composition based generative machine learning to discover new 2D materials, which greatly expand the space of 2D materials.
  \item we apply a template based element substitution based structure prediction approach to get the structures of hypothetical 2D materials and verify them using DFT formation energy calculations and phonon thermostability verification.
\end{itemize}

\section{Materials and Methods}

\subsection{2D materials discovery framework}

\begin{figure}[ht]
  \centering
  \includegraphics[width=\linewidth]{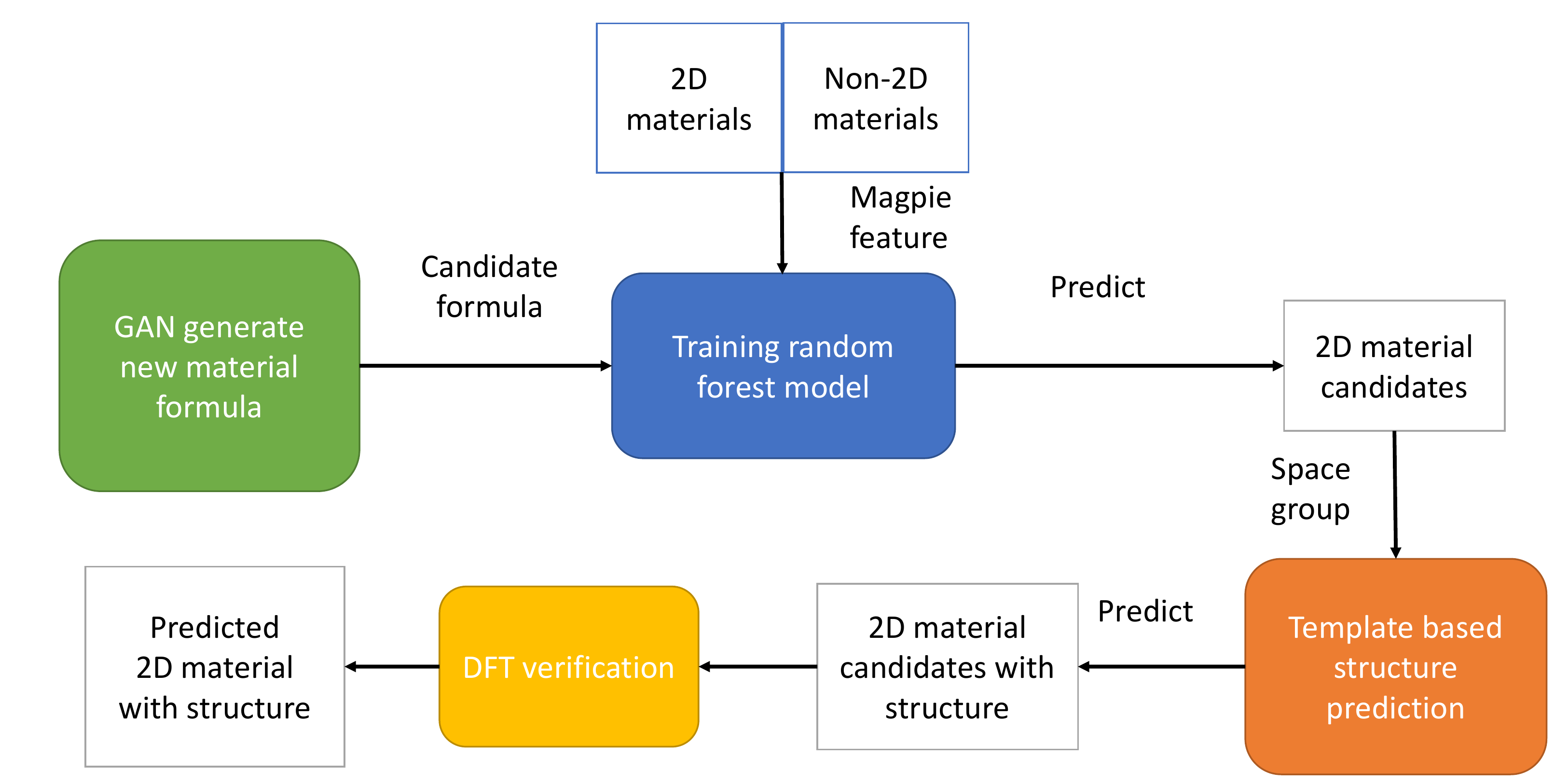}
  \caption{Framework for generation and prediction of 2D materials. It comprises four components. The green part: a GAN based composition generation module for breeding chemically valid materials. The blue part: a composition based random forest 2D materials classifier. The orange part: a template based element substitution structure predictor, as well as yellow part: DFT validation. }
  \label{framework}
\end{figure}

The schematic diagram of our 2D materials discovery framework includes the following four modules (Figure\ref{framework}): a GAN based hypothetical materials generator, a composition based 2D materials classifier, a template based structure predictor, and a DFT confirmation procedure. The hypothetical materials generator is trained with known inorganic materials in the Materials Project database to learn the composition rules of forming stable chemically valid materials compositions. Then, we use the generative module to breed a large number of hypothetical formulas (two million in our study). These formulas are then subjected to chemical validity tests including charge neutrality check and electronegativity check. After that, the remaining samples will be screened by the 2D materials classifier using composition alone. To verify the predicted 2D materials compositions, we apply template based element substitution to generate their hypothetical structures for a subset of 624 predicted 2D materials compositions. Using DFT calculations, the stability of these structures is calculated to verify the existence of these candidate 2D materials from which we identified twelve potentially stable materials.

\subsection{Generative deep learning for hypothetical inorganic materials}
In the material design research area, one core task is to explore chemical space for searching new materials. In our previous work, a generative machine learning model (MatGAN) \cite{dan2020generative} is designed to efficiently generate new hypothetical inorganic materials composition based on generative adversarial network (GAN) \cite{goodfellow2014generative}. There are two main tasks of MatGAN: one is how to suitably represent material composition; the other one is how to design the generative adversarial network for generating new materials.

\begin{figure}
  \centering
  \includegraphics[width=0.95\linewidth]{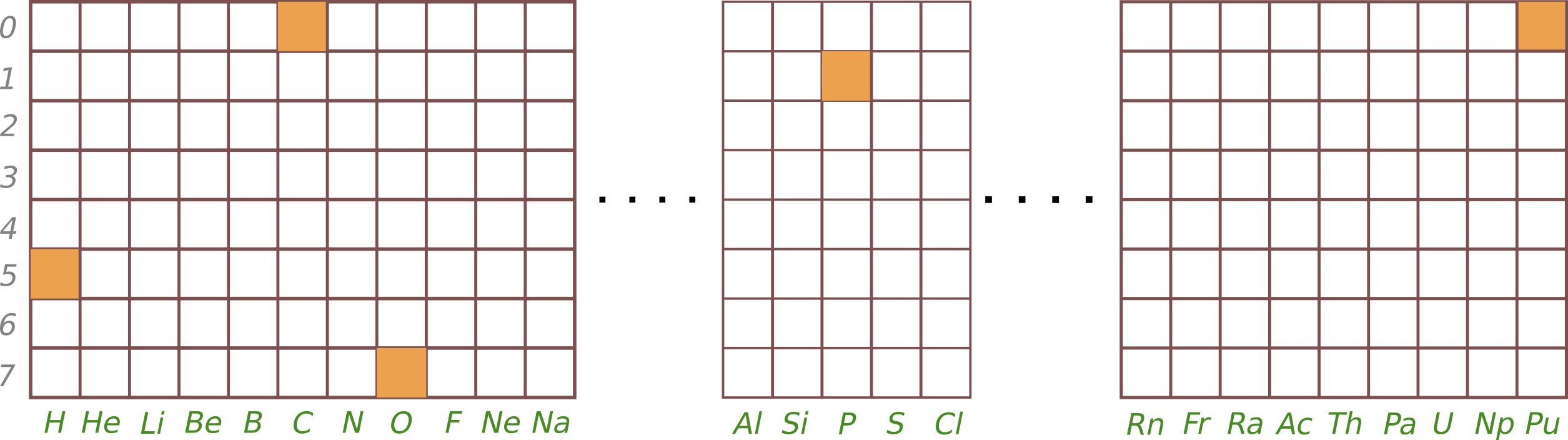}
  \caption{One hot representation of material composition PuP\textsubscript{2}H\textsubscript{6}CO\textsubscript{8}. Brown color indicates the atom number of corresponding element in the specific material.}
\label{fig:representation}
\end{figure}

When exploring the representation of inorganic materials, we found that there  totally are 85 elements in ICSD dataset; and there are no more than 8 atoms per element in any specific compound. Therefore, each material could be represented as a sparse matrix M of dimension $8\times 85$ with 0/1 cell values, where $M_{i,j}=1$ means the number of atoms of the element at column j is $i+1$. Figure \ref{fig:representation} shows the encoding matrix for PuP\textsubscript{2}H\textsubscript{6}CO\textsubscript{8}.

\begin{figure}
  \centering
  \includegraphics[width=0.8\linewidth]{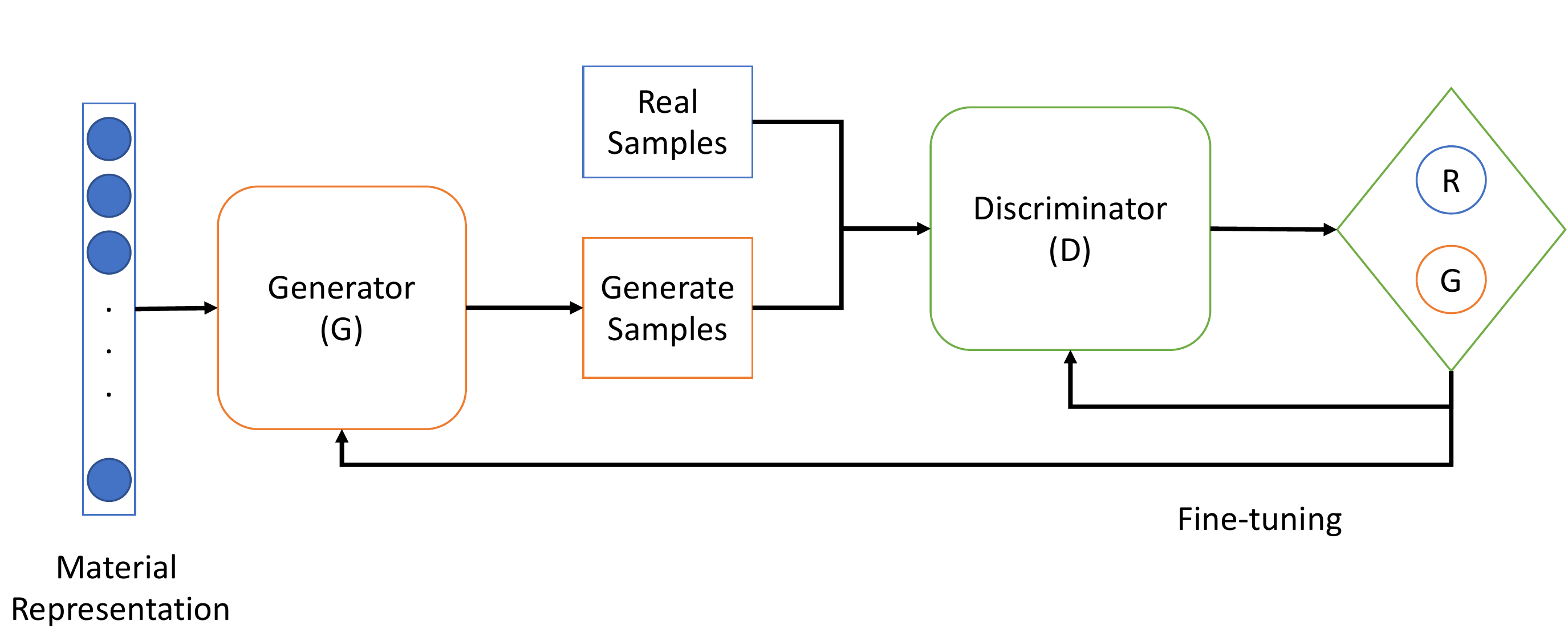}
  \caption{Architecture of MatGAN. Generator (G) learns from known material compositions to generate realistic samples while discriminator (D) learns to determine whether a sample is a real one or generated one. Alternative training of D and G will improve the performance of both G and D.}
\label{fig:matgan}
\end{figure}

The architecture of MatGAN is shown in Figure \ref{fig:matgan}. In this generative adversarial network training, a generator is trained from existing real material representations to generate new samples. Meanwhile, the discriminator tries to differentiate real samples from generated samples; as the feedback, the discrimination loss is then used to guide the training of the generator and the discriminator's parameters to reduce this difference. These two training processes are repeated until good performances of both the generator and the discriminator are achieved. The loss function is shown in the following equation:

\begin{equation}
\operatorname{Loss}=-\operatorname{Dice}=-\frac{2|A \cap B|}{|A|+|B|} \approx-\frac{2 \times A \bullet B}{\operatorname{Sum}(A)+\operatorname{Sum}(B)}
\end{equation}
where $A \cup B$  denotes the common elements of A and B, |g| represents the number of elements in a matrix, • denotes dot product, Sum(g) is the sum of all matrix elements. The dice coefficient essentially measures the overlap of two matrix samples with values ranging from 0 to 1, and 1 indicating perfect overlap.

We have generated 2,650,623 hypothetical materials compositions, and 1,940,209 of them satisfy both charge neutrality and electronegativity balance criteria. 

\subsection{Composition based classifiers for predicting 2D materials}

Predicting whether a material is 2D structure can be regarded as a binary classification problem. Our goal is to screen unknown 2D materials from MatGAN-generated materials by training a random forest classifier with verified 2D and non-2D materials, and then predicting the probability of being a 2D material of each new material. 

Specifically, we employ the Random Forest (RF) \cite{liaw2002classification} as the surrogate model for predicting the 2D probability given a material' Magpie composition features. Magpie feature set\cite{ward2018machine} is calculated using the matminer library \cite{ward2018matminer} which is a Python-based platform that facilitates data-driven methods for analyzing and predicting material properties by calculating a variety of descriptors from material compositions or crystal structures. Magpie feature set summarizes 132 features about material composition information, such as atomic number in the material, range of melting temperature, mean absolute deviation of valence. Those 132 features of each material will be calculated and used in our random forest model training.

RF is a kind of supervised bagging ensemble learning algorithm. The idea behind random forests is to exploit the wisdom of the group. RF builds many decision trees in a random way with low correlation among them. After building the forest, when a new sample needs to be classified, each decision tree makes a judgment separately to vote which category the sample belongs to. Random forest improves the prediction accuracy without significantly increasing the amount of computation, and it is relatively robust to unbalanced data. 

In the data preparation stage, we first collect known 2D and non-2D materials, as well as MatGAN generated new materials. Then, we calculate the Magpie features for all of them. In training the random forest model, known 2D materials and non-2D materials are treated as positive and negative samples to train the RF classifier with 10-fold cross-validation. The RF hyper-parameters are tuned to achieve good prediction performance with detailed settings explained in Section 3.2.2. Afterward, the trained RF model is utilized to predict the labels and probability scores of 2D for generated hypothetical new materials.

\subsection{Template based structure prediction}
\begin{figure}[ht]
  \centering
  \includegraphics[width=\linewidth]{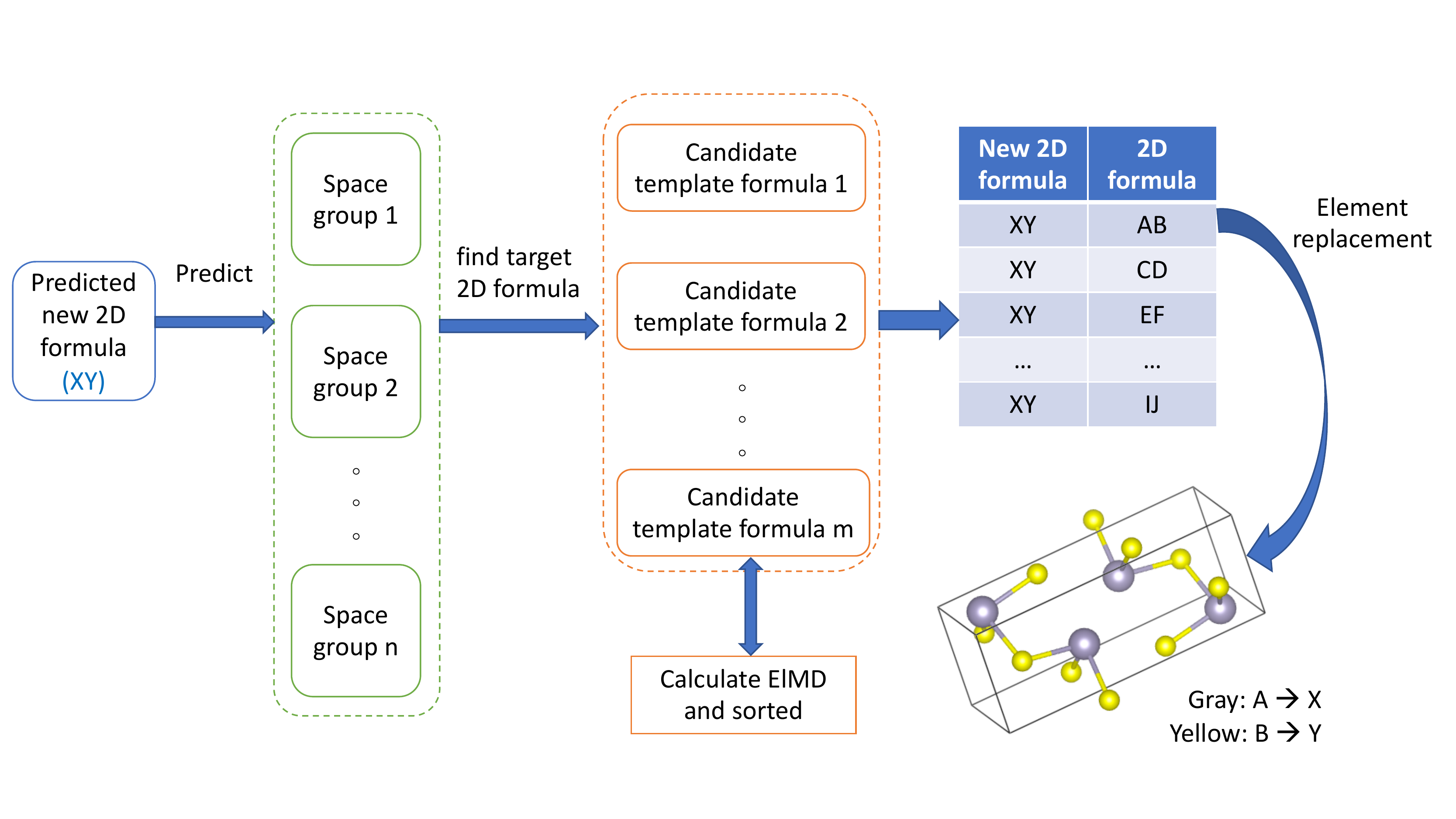}
  \caption{The framework of template based structure prediction. The main parts are: predicting new formula's space group; finding candidate template formulas according to the same element number and the same space group; calculating and sorting ElMD between the new formula and template formulas; selecting top n templates to do element substitution to get the new formula's structure. }
  \label{fig:template}
\end{figure}

Although we have predicted the 2D probability scores of the hypothetical new materials generated by our MATGAN algorithm and the candidates with their probability scores greater than 95\% are likely to be 2D materials, it is not enough to verify their existence by DFT calculation of their formation energy or phonon calculation based stability check. However, crystal structure prediction of complex compositions using current ab initio crystal structure prediction algorithms are not feasible \cite{oganov2019structure}. To address this issue, we propose to use the template based or element substitution based structure prediction method, which is shown in Figure \ref{fig:template}.

Firstly, for each predicted 2D formula, we use the Crystal Structure Prediction Network (CRYSPNet) \cite{liang2020cryspnet} tool to predict its space group that the formula most likely belongs to. This method consists of many neural network models to predict the material's space group, Bravais lattice, and lattice constants. As CRYSPNet only needs chemical composition information as input, we use it to estimate the top 3 potential space groups for each new hypothetical 2D material.

Next, we try to find similar template materials from known 2D materials in the 2dMatpedia database. Specifically, For each new 2D formula with three potential space groups, we search the target 2D material that has the same number of elements and the same space group. However, one formula may lead to many potential target 2D material template. To identify the most similar template material, we use Element Movers Distance (ElMD) \cite{hargreaves2020earth} to calculate and sort the similarity between the candidate material and potential template materials. ElMD is a similarity measure for chemical compositions, which is measured by the minimal amount of work taken to transform one distribution of elements to another along the modified Pettifor scale.

Finally, we select the top 10 most similar known 2D materials as the structure templates according to the ElMD values. New 2D material's structures could be then predicted by one to one element substitution from those pairs. For example, as shown in Figure \ref{fig:template}, the structure of XY is predicted from the structure of AB by using X to replace A (gray atom) and Y to replace B (yellow atom).

\subsection{DFT calculation for verification}

The density functional theory (DFT) calculations were performed based on the  Vienna \textit{ab initio} simulation package (VASP) \cite{Vasp1,Vasp2,Vasp3,Vasp4}. The electron-ion interactions were considered by using the projected augmented wave (PAW) method \cite{PAW1, PAW2}. The energy cutoff value was set as 500 eV. The generalized gradient approximation (GGA) based on the Perdew-Burke-Ernzerhof (PBE) pseudopotentials \cite{GGA1, GGA2} represented the exchange-correlation potentials. The energy convergence criterion was set as 10$^{-7}$ eV, while the force convergence criterion of the ionic steps was considered as 10$^{-2}$ eV/{\AA}.  The $\Gamma$-centered  Monkhorst-Pack $k$-meshes were considered to perform the Brillouin zone integration for the unit cells. Formation energy per atom of a material were calculated based on  Eq.\ref{eq:form}. Here, $E[\mathrm{Material}]$ is the total energy per unit formula of corresponding material, $E[A_i]$ is the energy of $i^\mathrm{th}$ element of the material and N indicates the total number of atoms in a unit formula of the material. 

\begin{equation}
    E_{\mathrm{form}} =\frac{1}{N}(E[\mathrm{Material}] - \sum_i E[A_i])
    \label{eq:form}
\end{equation}

\section{Experiments}
\subsection{2D materials Dataset}
All the 2D materials are collected from 2DMatPedia\cite{zhou20192dmatpedia}, an open computational database of two-dimensional materials, which is constructed by a topology based screening algorithm and element substitutions. There are 6,351 2D materials in total, they are regarded as positive training samples in our work. We also collect all existing materials from the materials project database with 126,356 materials in total. After removing known 2D materials, there are 115,498 negative samples. We use the MatGAN model to generate 2,650,264 new materials as candidates for 2D materials prediction. 

\begin{table}
\begin{center}
\caption{Datasets }
\begin{tabular}{ccc}
\hline
Dataset         & Amount    & Role                     \\ \hline
2dMaterials     & 6,351     & positive training sample \\
MaterialProject & 126,356   & negative training sample (exclude 2D materials) \\
ICSD\_2M        & 2,650,624 & potential new material   \\
V2DB            & 294,077   & comparative dataset      \\ \hline
\end{tabular}
\label{table:dataset}
\end{center}
\end{table}


\subsection{Results}

\subsubsection{Generation of candidate inorganic materials}

Trained with 291,840 inorganic materials compositions in Material Project database, a generative deep learning model (MATGAN) is used to generate 2,650,623 new compositions, and then charge neutrality and balanced electronegativity are used to screen out 1,947,792 formulas, among which 1,940,209 are not in the training set. In order to better display the generated materials distribution information, we draw a histogram based on the element number, and another line chart to show the frequencies of 112 elements.

\begin{figure}[ht]
  \centering
  \includegraphics[height=6cm]{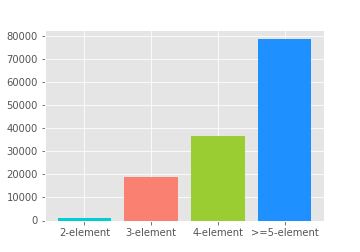}
  \caption{Generated materials distribution information. }
  \label{fig:element number}
\end{figure}

The distribution of generated materials based on element numbers is shown in Figure \ref{fig:element number}. The number of 2-element, 3-element, 4-element, >=5=element are 1217, 18946, 36827, 78639, respectively. There are two reasons to explain why binary materials account for the least proportion: one is the diversity of the combination of two elements is much less than that of five elements. The other is many binary materials have already been discovered. The generated 2-element materials play an important role in subsequent research because most known 2D materials are binary materials.

\begin{figure}[ht]
  \centering
  \includegraphics[width=1\linewidth]{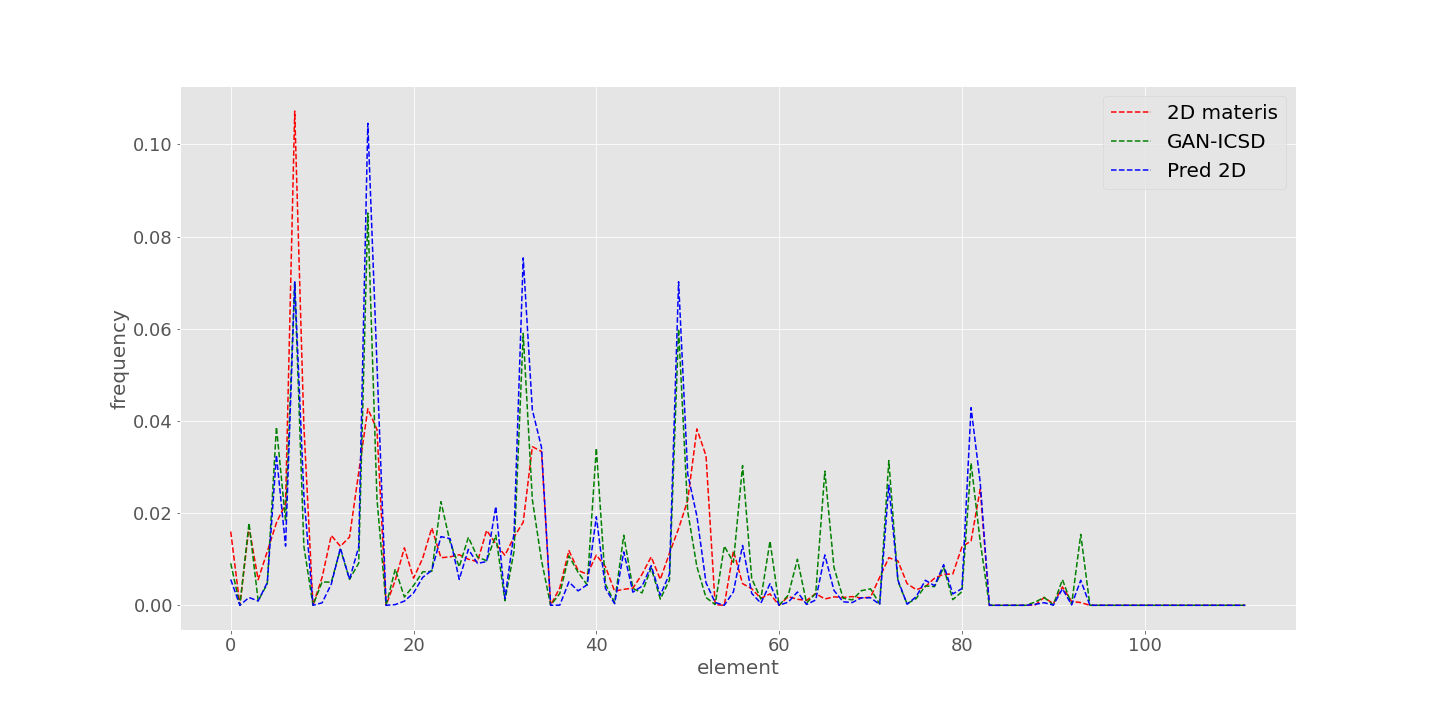}
  \caption{Element frequency distribution. The abscissa represents 112 chemical elements arranged according to the atomic number from hydrogen (H) to Copernicium (Cn), vertical axis indicates the frequency of each element. The red curve denotes the distribution information of 2dMaterials dataset. The green curve shows our generated candidate materials. The blue line represents our predicted 2D materials.}
  \label{fig:frequency}
\end{figure}

Figure \ref{fig:frequency} shows the frequency distribution of 112 elements ordered by atomic number in three datasets: generated ICSD-2M candidate materials by MatGAN, the published 2D material dataset and the predicted 2D materials. From these three curves, we can see that the top 5 crests positions basically overlap, the number ranges are 7-9, 15-17, 32-35, 50-53, and 80-83. It proves that the space of candidate materials generated by MatGAN is consistent with the real 2D materials. Furthermore, it provides a solid candidate range for the following 2D new material prediction.

\subsubsection{Performance of the 2D materials classifier}
The hyper-parameter configuration for training 2D random forest classifier is set as follows: we set the maximum tree depth to be 20 and the number of decision trees as 250. There are 6351 2D material samples and 15,959 non-2D samples. In order to mitigate the imbalanced positive and negative samples, we randomly select 1.5 times the number of positive samples as negative samples. Besides, the class weight parameter is set to be balanced. With these settings tuned per feature iteration, we train the RF 2D materials prediction models and evaluate their performance. Our algorithm is implemented using the Scikit-Learn library in Python 3.6.

To evaluate the prediction performance of our model, precision, recall, accuracy, F1 score, and receiver operating characteristic area under the curve ROC are used as performance metrics. ROC is plotted with TPR and FPR as the vertical and horizontal axes under different threshold settings. 

\begin{equation}
  TPR =  \frac{TP}{TP + FN}
  \label{TPR}
\end{equation}

\begin{equation}
  FPR =  \frac{FP}{FP + TN}
  \label{FPR}
\end{equation}

\begin{figure}[!ht]
  \centering
  \begin{subfigure}{.45\textwidth}
    \includegraphics[width=7.8cm, height=6cm]{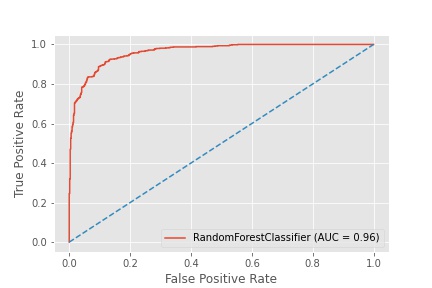}
    \caption{ROC curve of the 2D materials RF classifier}
    \vspace{3pt}
  \end{subfigure}
  \begin{subfigure}{.45\textwidth}
    \includegraphics[width=7.8cm, height=6cm]{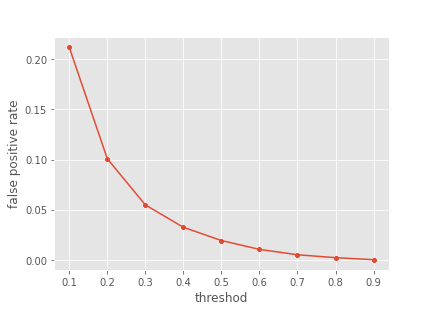}
    \caption{False-positive curve of the 2D materials RF classifier.}
    \vspace{3pt}
  \end{subfigure}

  \caption{performance of random forest classifier.}
  \label{fig:result}
\end{figure}

The accuracy, precision, recall, and F1-measure of the RF classifier with 10 fold cross-validation is 88.97\%, 88.98\%, 88.96\%, and 88.96\%. Figure \ref{fig:result}(a) shows the ROC curve of the classifier with an AUC score reaching 96\%. 

We also use a series of thresholds to differentiate 2D and non-2D materials to evaluate the performance of the RF classifier, as shown in Figure \ref{fig:result}(b). The abscissa represents the  the predicted probability threshold to declare a 2D materials when its probability is higher than this threshold value; the y coordinate indicates the corresponding false-positive rate. As the threshold increases, the higher the probability score is required to be judged as a 2D material leading to lower false-positive rate.

\subsubsection{Finding new 2D materials using our framework}

To identify interesting hypothetical new 2D materials, we applied our RF-based 2D materials classifiation model to screen the 2.6 million hypothetical materials generated by our Generative Adversarial Network (GAN) based on new materials composition generator \cite{dan2020generative}. After predicting the probability of each candidate belonging to 2D materials, we sort them by the probability scores. The statistics of the predicted 2D materials with different probability thresholds are shown in Table\ref{table:predict 2d material}. With a stringent probability threshold of 0.95, our algorithm has identified 1,485 hypothetical 2D material formulas with 266 binary, 361 ternary, 327 quartenary candidates. When the threshold is lowered to 0.9, the number of candidate 2D formulas increases to 5,034 or to 18,451 with threshold of 0.8. 

\begin{figure}
  \centering
  \begin{subfigure}{.45\textwidth}
    \includegraphics[width=\textwidth]{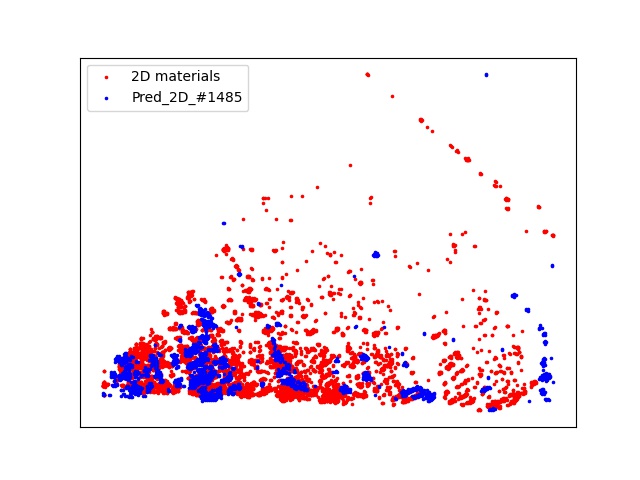}
    \caption{New 2D materials (Probability > 0.95) vs. known ones }
    \vspace{3pt}
  \end{subfigure}
  \begin{subfigure}{.45\textwidth}
    \includegraphics[width=\textwidth]{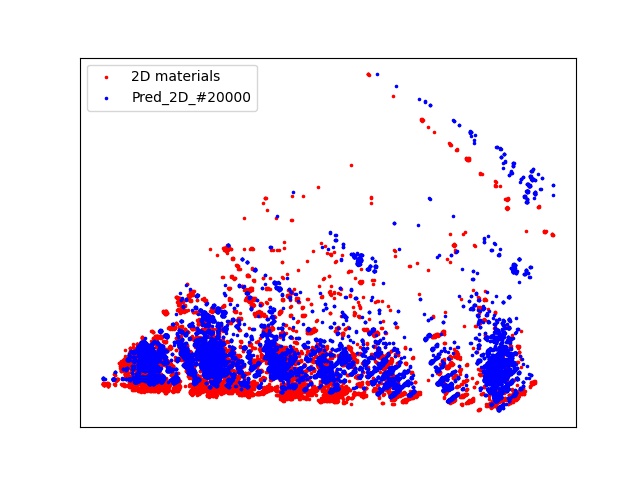}
    \caption{20,000 2D materials in the whole MP composition space }
    \vspace{3pt}
  \end{subfigure}

  \begin{subfigure}{.45\textwidth}
    \includegraphics[width=\textwidth]{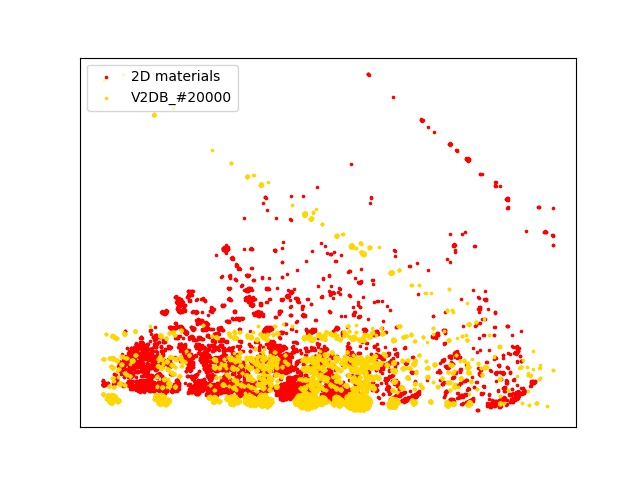}
    \caption{20,000 2D materials in the V2DB }
    \vspace{3pt}
  \end{subfigure}
  \begin{subfigure}{.45\textwidth}
    \includegraphics[width=\textwidth]{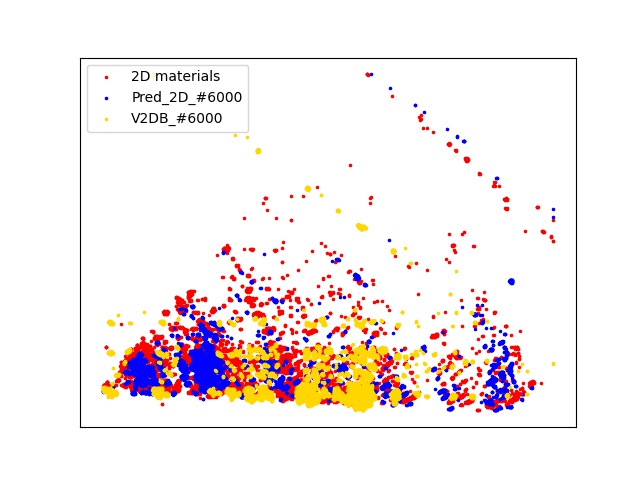}
    \caption{6,000 2D materials in MP and V2DB vs. known ones}
    \vspace{3pt}
  \end{subfigure}
  
  \caption{Distribution of the new and old 2D materials using t-sne visualization. Red ones are known 2D materials, blue ones are our predicted 2D materials while yellow ones are predicted 2D materials from V2DB. Figure (a) shows 1485 points with the highest 2D probability scores. (b) and (c) display distribution of 20,000 predicted samples from our model and from V2DB respectively. For better visual comparison, 6,000 predicted 2D materials from our model and from V2DB are shown together in (d).}
  \label{fig:2d_distribution}
\end{figure}

To demonstrate how the newly predicted 2D materials are distributed in the composition space, we apply t-sne dimension reduction tool \cite{maaten2008visualizing} to map the Magpie features of the 6351 2D materials in the 2Dmatpedia database and the predicted 2D materials, and then plot their distribution in Figure \ref{fig:2d_distribution}.  In Figure \ref{fig:2d_distribution}(a), We apply the same dimension reduction transformation to both the training set and the newly predicted 2D materials and visualize their distribution where red points are training samples, and blue points are the 1485 predicted 2D materials with the highest probability scores. Similarly, top 20,0000 predicted 2D materials are drawn as blue points in Figure \ref{fig:2d_distribution}(b) together with known 2D materials used for training. We found that in Figure \ref{fig:2d_distribution}(a) the majority of blue points are located in the dense red point areas in the bottom left corner (which can be seen also from Figure \ref{fig:2d_distribution}(c)), indicating that our predicted 2D materials have similar composition distribution with regard to known 2D materials. Figure \ref{fig:2d_distribution}(b) further confirms this composition distribution match, in which we find that the blue points in general only appear in areas with red points. The areas with sparse red points also contain few blue points. Figure \ref{fig:2d_distribution}(c) shows the distribution of 20,000 hypothetical 2D materials in the V2DB dataset against to the known 2D materials. It is found these candidate 2D materials have different composition distribution as regard to the known 2D materials: many yellow points appear in areas without red points. Quite many yellow points reach out of the boundary defined by the red points. For better comparison, top 6000 predicted samples by our method and the V2DB are drawn in Figure \ref{fig:2d_distribution}(d) together with 6351 known 2D materials. It can be seen that the majority of the overlapped blue and yellow points (117 as shown in Table\ref{table:overlap predic 2d}) are located in the lower-left corner, which means there are new 2D materials that are jointly predicted by both methods.

To screen out top candidate materials, we use the Roost algorithm \cite{goodall2019predicting} for formation energy prediction, which is a graph network based machine learning model for materials property prediction using only composition information. After predicting the formation energies of all candidates, we draw a histogram of formation energy distribution, as shown in Figure \ref{fig:frequency}. Furthermore, we filter out those with probability scores greater than 0.95 and then sort them by the formation energy in ascending order and pick the top 40 candidates with 2, 3, and 4 elements respectively. The results are in Table \ref{table:predic 2d}. In our supplementary file, we have listed 1000 top hypothetical 2D materials with high 2D probability scores and low predicted formation energies.

Furthermore, we analyze the 2DMatPedia dataset, 2-element materials occupy 65\%, 3-element materials, and 4-element materials account for 25\% and 9\%, therefore, from the perspective of the probability distribution, our prediction is meaningful. We also find that the predicted 2D probabilities of 2-element materials are in general higher than those of 3-element materials and 4-element materials, corresponding to the fact that the majority of known 2D materials (65\%) are binary materials. We also count the number of our predicted new 2D materials with 2D probability greater than 0.5 that overlap with those in V2DB and 117 hypothetical 2D materials are found to be predicted by both methods. Table \ref{table:overlap predic 2d} shows the overlapped candidate 2D materials in five parts according to their 2D probability scores. It is found the overlapped materials with 2D probability greater than 0.8 account for nearly 50\% of all overlapped candidates.

\begin{table}
\begin{center}
\caption{Statistics of predicted 2D materials }
\label{table:predict 2d material}

\begin{tabular}{cccccc}
\hline
2D Probability & \# of Predicted new 2D formula & \# 2 element & \# 3 element & \# 4 element & \# \textgreater{}= 5 element \\ \hline
0.95         & 1,485                     & 266          & 861          & 327          & 31                           \\
0.9          & 5,034                     & 439          & 2,617        & 1,695        & 283                          \\
0.8          & 18,451                    & 729          & 8,123        & 7,316        & 2,283                        \\
0.7          & 48,592                    & 942          & 16,827       & 21,430       & 9,393                        \\
0.6          & 119,489                   & 1,146        & 28,172       & 51,998       & 38,173                       \\
0.5          & 267,489                   & 1,340        & 40,382       & 99,943       & 125,824                      \\ \hline

\end{tabular}
\end{center}
\end{table}

\begin{figure}[ht]
  \centering
  \includegraphics[width=0.7\linewidth]{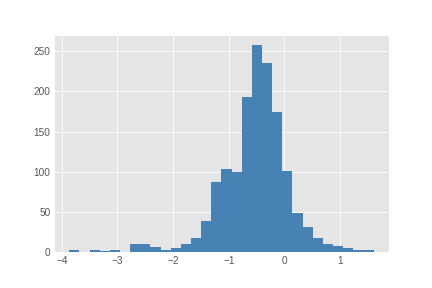}
  \caption{Formation energy distribution of predicted 2D materials.}
  \label{fig:histogramformation}
\end{figure}

\begin{table}[!ht]
\centering
\caption{Predicted hypothetical 2D materials that overlap with V2DB}
\begin{tabular}{cccccccccc}
\hline
Formula & Probability         & Formula & Probability         & Formula & Probability         & Formula & Probability         & Formula & Probability         \\
        & \textgreater{}0.9 &         & \textgreater{}0.8 &         & \textgreater{}0.7 &         & \textgreater{}0.6 &         & \textgreater{}0.5 \\ \hline
AgS     & 0.9999            & ZrClS   & 0.8997            & TiSeCl  & 0.7913            & VISe    & 0.6879            & NiF     & 0.5998            \\
AlS     & 0.9982            & AlIS    & 0.8926            & ZrSeF   & 0.7900            & NiSeCl  & 0.6830            & ZrTeN   & 0.5993            \\
ScI     & 0.9840            & VClF    & 0.8917            & SbSeO   & 0.7856            & CoBrF   & 0.6797            & MnSeS   & 0.5872            \\
InTeS   & 0.9753            & FeClS   & 0.8909            & ZrTeSe  & 0.7802            & NiSF    & 0.6794            & SrSF    & 0.5840            \\
SnTeSe  & 0.9748            & PbTeS   & 0.8884            & MnBr    & 0.7793            & RhSeS   & 0.6759            & YSeS    & 0.5709            \\
SnTeS   & 0.9737            & GeSF    & 0.8852            & MnIS    & 0.7782            & CoTeS   & 0.6740            & NbSeO   & 0.5669            \\
PbTeSe  & 0.9465            & BiSF    & 0.8830            & CuTeO   & 0.7647            & CrSeO   & 0.6732            & CoF     & 0.5638            \\
SbSeF   & 0.9421            & TiClF   & 0.8800            & CoCl    & 0.7602            & CoSF    & 0.6714            & NbTeO   & 0.5636            \\
SbClS   & 0.9418            & ZnBrS   & 0.8777            & GeSO    & 0.7588            & RhTeS   & 0.6710            & TaSO    & 0.5604            \\
GeClS   & 0.9409            & MnClS   & 0.8757            & ZnSF    & 0.7579            & MnBrN   & 0.6697            & TaTeS   & 0.5506            \\
AsClS   & 0.9374            & SnSeO   & 0.8727            & ZnSO    & 0.7579            & AlSeO   & 0.6609            & NiSO    & 0.5474            \\
SbSF    & 0.9224            & ZnClS   & 0.8725            & AgSO    & 0.7557            & PbTeO   & 0.6540            & NiSeO   & 0.5419            \\
NbSe    & 0.9221            & AgIS    & 0.8670            & CuSeF   & 0.7546            & MnSeO   & 0.6473            & YTeS    & 0.5249            \\
AlSeS   & 0.9197            & PbSeO   & 0.8535            & MnSF    & 0.7518            & VSeS    & 0.6451            & TiSeN   & 0.5186            \\
SnSO    & 0.9181            & AsSF    & 0.8532            & MnSeF   & 0.7466            & NbSeF   & 0.6367            & SrClS   & 0.5120            \\
BiTeS   & 0.9102            & AgTeS   & 0.8524            & AsSO    & 0.7422            & RuTeS   & 0.6310            &         &                   \\
AlTeS   & 0.9060            & PbSO    & 0.8494            & AgSeO   & 0.7406            & TaSeO   & 0.6217            &         &                   \\
        &                   & YClS    & 0.8471            & ZrSeS   & 0.7392            & FeSeF   & 0.6190            &         &                   \\
        &                   & BiTeF   & 0.8462            & CoClF   & 0.7362            & TaTeSe  & 0.6077            &         &                   \\
        &                   & NiClS   & 0.8457            & PdFO    & 0.7319            & CrTeO   & 0.6066            &         &                   \\
        &                   & YTeF    & 0.8457            & ZrTeO   & 0.7317            & ZrSeN   & 0.6014            &         &                   \\
        &                   & CoClS   & 0.8433            & MnI     & 0.7312            &         &                   &         &                   \\
        &                   & VClS    & 0.8397            & CuSF    & 0.7245            &         &                   &         &                   \\
        &                   & VSF     & 0.8395            & AlSO    & 0.7239            &         &                   &         &                   \\
        &                   & AlTeCl  & 0.8285            & TiTeS   & 0.7230            &         &                   &         &                   \\
        &                   & NiBr    & 0.8254            & NiClO   & 0.7077            &         &                   &         &                   \\
        &                   & SbSeS   & 0.8220            & NiTeCl  & 0.7056            &         &                   &         &                   \\
        &                   & AgSeS   & 0.8217            & NbSO    & 0.7005            &         &                   &         &                   \\
        &                   & BiSO    & 0.8201            &         &                   &         &                   &         &                   \\
        &                   & ZnClF   & 0.8176            &         &                   &         &                   &         &                   \\
        &                   & TiSeF   & 0.8155            &         &                   &         &                   &         &                   \\
        &                   & MnCl    & 0.8077            &         &                   &         &                   &         &                   \\
        &                   & FeI     & 0.8074            &         &                   &         &                   &         &                   \\
        &                   & AlSeF   & 0.8017            &         &                   &         &                   &         &                   \\
        &                   & AlTeF   & 0.8015            &         &                   &         &                   &         &                   \\
        &                   & CoI     & 0.8002            &         &                   &         &                   &         &                   \\ \hline
\end{tabular}

\label{table:overlap predic 2d}
\end{table}

\begin{table}[!h]
\centering
\caption{Hypothetical 2D materials sorted by predicted formation energy(only top 120 are listed here)}
\begin{tabular}{ccccccccc}
\hline
\multicolumn{3}{c}{2   Elements}                                                    & \multicolumn{3}{c}{3 Elements}                                                      & \multicolumn{3}{c}{4 Elements}                                                          \\
Formula & Probability & \begin{tabular}[c]{@{}c@{}}Formation\\  Energy\end{tabular} & Formula & Probability & \begin{tabular}[c]{@{}c@{}}Formation \\ Energy\end{tabular} & Formula     & Probability & \begin{tabular}[c]{@{}c@{}}Formation \\ Energy\end{tabular} \\ \hline
ZrF3    & 0.9876      & -3.8661                                                     & ScYF3   & 0.9760      & -3.3429                                                     & NbMoCl5S2   & 0.9757      & -1.5188                                                     \\
YF2     & 0.9880      & -3.7549                                                     & NbMoF6  & 0.9720      & -3.0069                                                     & CrNbMoCl8   & 0.9638      & -1.4547                                                     \\
TaF4    & 0.9990      & -3.4457                                                     & ZnTaF5  & 0.9755      & -2.9848                                                     & CrNbRuCl8   & 0.9598      & -1.3912                                                     \\
SiF3    & 0.9607      & -3.3363                                                     & NbRuF7  & 0.9665      & -2.7521                                                     & NbRuCl5S2   & 0.9753      & -1.3857                                                     \\
ZrF2    & 0.9997      & -3.1413                                                     & InSnF5  & 0.9519      & -2.7437                                                     & NbRuCl4S2   & 0.9712      & -1.3826                                                     \\
ScF     & 0.9760      & -2.7647                                                     & TaIrF7  & 0.9794      & -2.7379                                                     & NbRuCl6S    & 0.9673      & -1.3719                                                     \\
YF      & 0.9640      & -2.747                                                      & NbRuF6  & 0.9708      & -2.7165                                                     & NbRuCl6S2   & 0.9633      & -1.3688                                                     \\
GeF3    & 0.9705      & -2.6997                                                     & TaWO5   & 0.9732      & -2.6845                                                     & CrMoCl5S    & 0.9517      & -1.3163                                                     \\
NbF2    & 0.9671      & -2.5717                                                     & ScYCl6  & 0.9677      & -2.6222                                                     & NbMoRuCl6   & 0.9676      & -1.2988                                                     \\
YCl2    & 0.9919      & -2.4657                                                     & NbMoO5  & 0.9816      & -2.6101                                                     & Ga3AsCl6S2  & 0.9514      & -1.2548                                                     \\
ZrF     & 0.9720      & -2.4356                                                     & TaWF5   & 0.9760      & -2.5844                                                     & InSn3SeCl6  & 0.9513      & -1.2127                                                     \\
WF3     & 0.9999      & -2.4182                                                     & TaIrF5  & 0.9511      & -2.5832                                                     & SnSbAsCl8   & 0.9593      & -1.205                                                      \\
TaF2    & 0.9880      & -2.4126                                                     & ScYCl4  & 0.9679      & -2.527                                                      & MoRuCl5S2   & 0.9633      & -1.171                                                      \\
AlF     & 0.9999      & -2.4107                                                     & YZrCl6  & 0.9598      & -2.4481                                                     & Sn2AsCl6S   & 0.9550      & -1.1705                                                     \\
ScCl2   & 0.9799      & -2.338                                                      & ScZrCl6 & 0.9518      & -2.3783                                                     & InSn3Cl4S2  & 0.9907      & -1.1704                                                     \\
GaF     & 0.9995      & -2.0765                                                     & MoRuF6  & 0.9708      & -2.352                                                      & MoRuCl6S2   & 0.9633      & -1.1499                                                     \\
FB      & 0.9880      & -2.0746                                                     & TaOsF5  & 0.9640      & -2.3308                                                     & Sb2BrAsCl8  & 0.9671      & -1.1406                                                     \\
F3C     & 0.9800      & -2.0634                                                     & ScTiCl4 & 0.9600      & -2.2386                                                     & InSnCl2S2   & 0.9745      & -1.1381                                                     \\
InF     & 0.9880      & -1.9397                                                     & VTc2F7  & 0.9661      & -2.2366                                                     & InSnCl4S2   & 0.9709      & -1.1229                                                     \\
Tc2O5   & 0.9650      & -1.9029                                                     & VZrCl6  & 0.9560      & -1.9388                                                     & SnSbAsCl4   & 0.9514      & -1.1043                                                     \\
ErBr3   & 0.9674      & -1.8561                                                     & YLaBr6  & 0.9547      & -1.9262                                                     & Sn2AsCl4S   & 0.9628      & -1.0983                                                     \\
OsF3    & 0.9837      & -1.7279                                                     & TbDyBr6 & 0.9729      & -1.8741                                                     & Sn2AsCl8S2  & 0.9627      & -1.0955                                                     \\
NbCl3   & 0.9959      & -1.7273                                                     & TiVCl6  & 0.9519      & -1.8074                                                     & Sn3SeCl6S2  & 0.9509      & -1.0928                                                     \\
S2O5    & 0.9859      & -1.7237                                                     & Nb3Cl8S & 0.9919      & -1.7429                                                     & Sn2AsCl7S2  & 0.9587      & -1.0901                                                     \\
NbCl2   & 0.9799      & -1.7137                                                     & NbCl2S  & 0.9599      & -1.6969                                                     & SnPb2Br2Cl2 & 0.9573      & -1.0892                                                     \\
OB      & 0.9600      & -1.7046                                                     & AlClS   & 0.9596      & -1.6887                                                     & Sn2AsCl6S2  & 0.959       & -1.0838                                                     \\
TaCl2   & 0.9950      & -1.6658                                                     & TaRe2F6 & 0.9640      & -1.6724                                                     & InSn2Cl2S2  & 0.9755      & -1.0789                                                     \\
RuF2    & 0.9864      & -1.5312                                                     & CrNbCl7 & 0.9677      & -1.6054                                                     & Sn2As2Cl6S  & 0.9511      & -1.0705                                                     \\
SF      & 0.9798      & -1.4954                                                     & CrNbCl6 & 0.9758      & -1.602                                                      & InSnClS2    & 0.9820      & -1.0414                                                     \\
MnCl3   & 0.9997      & -1.4641                                                     & CrNbCl8 & 0.9637      & -1.5821                                                     & Sn2AsCl4S2  & 0.9628      & -1.0387                                                     \\ 
CrCl4   & 0.9838      & -1.4529                                                     & TiYBr6  & 0.9538      & -1.5738                                                     & CuRuAgCl6   & 0.9629      & -1.0133                                                     \\
SeF     & 0.9795      & -1.436                                                      & SnPbCl6 & 0.9577      & -1.5723                                                     & Sn2AsCl6S3  & 0.9587      & -1.011                                                      \\
O3C2    & 0.992       & -1.3673                                                     & VMnCl4  & 0.96        & -1.5719                                                     & Sn2SeCl6S2  & 0.9509      & -1.0109                                                     \\
GeCl3   & 0.9988      & -1.3663                                                     & CrNbCl5 & 0.9598      & -1.5622                                                     & Sn2SeCl5S2  & 0.9549      & -1.0048                                                     \\
YS3     & 0.9618      & -1.3661                                                     & Te2SO5  & 0.9508      & -1.5303                                                     & AsGeCl6S    & 0.9590      & -1.0045                                                     \\
V2S3    & 0.9661      & -1.3311                                                     & InSnCl6 & 0.9795      & -1.5166                                                     & AsGe2Cl6S2  & 0.9636      & -0.9982                                                     \\
TiCl    & 0.9599      & -1.3269                                                     & SnTlCl5 & 0.9507      & -1.5121                                                     & Sn2As2Cl6S3 & 0.9547      & -0.9832                                                     \\
HoS3    & 0.9755      & -1.3082                                                     & TaWCl6  & 0.9794      & -1.5067                                                     & Sn2SeCl4S2  & 0.9549      & -0.9821                                                     \\
SmS3    & 0.9555      & -1.3041                                                     & NbMoCl6 & 0.9839      & -1.5054                                                     & Sn2AsCl6S4  & 0.9590      & -0.977                                                      \\
YI2     & 0.9999      & -1.2846                                                     & Cl6SSi2 & 0.9877      & -1.5012                                                     & SnPb2Se2Cl2 & 0.9538      & -0.9457 \\
\hline
\end{tabular}

\label{table:predic 2d}
\end{table}

\subsubsection{Structure prediction and verification }
\begin{figure}[!ht]
  \centering
  \begin{subfigure}{.45\textwidth}
    \includegraphics[width=7.8cm, height=6cm]{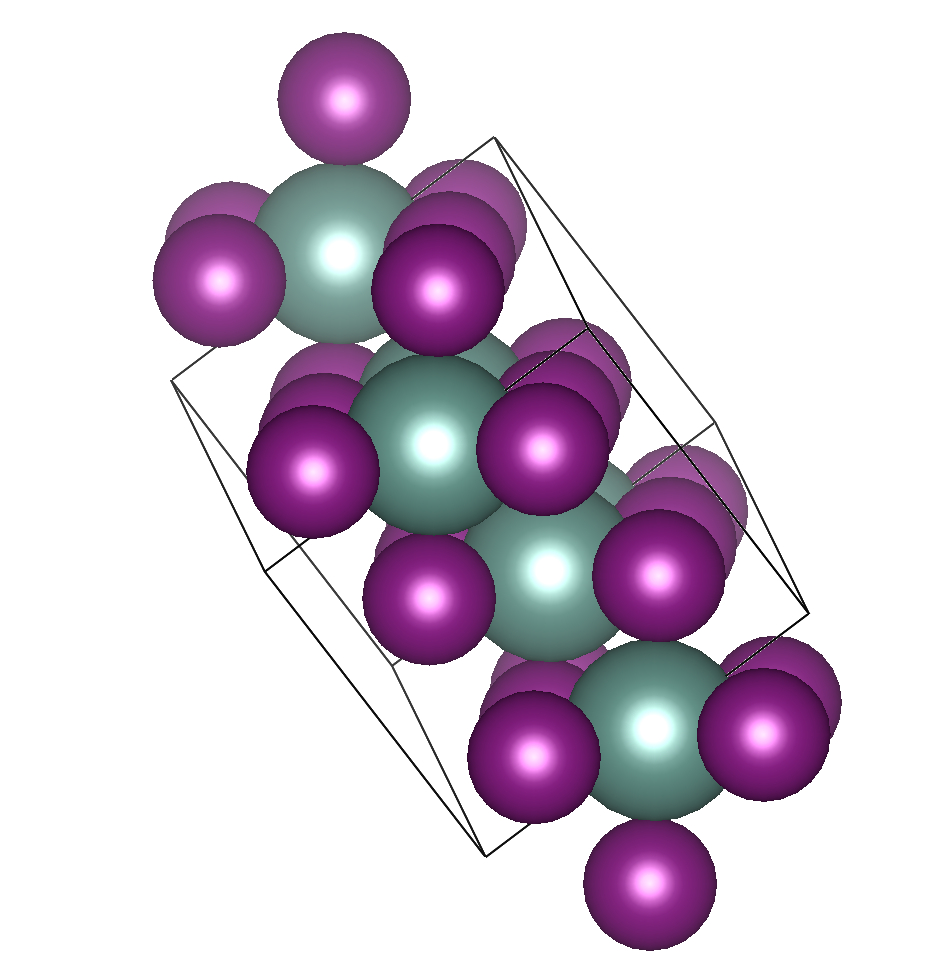}
    \caption{2D material YI2 by element substitution with formation energy -1.26 eV}
    \vspace{3pt}
  \end{subfigure}
  \begin{subfigure}{.45\textwidth}
    \includegraphics[width=7.8cm, height=6cm]{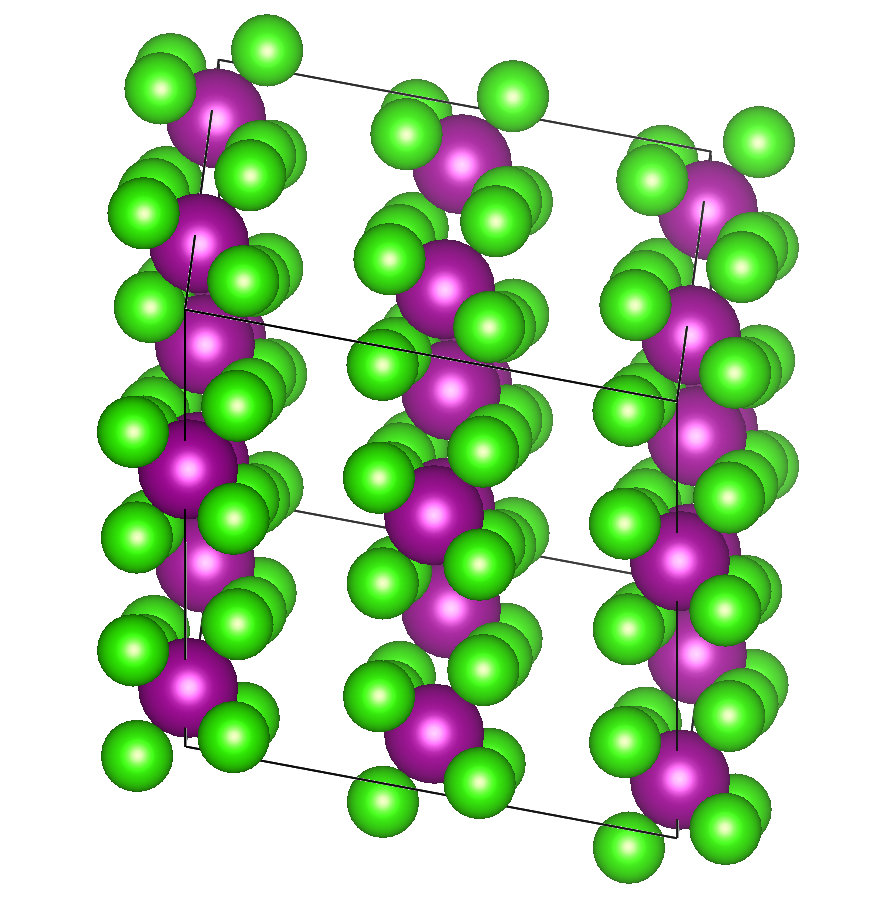}
    \caption{2D materials MnCl3 by element substitution with formation energy -1.45 eV}
    \vspace{3pt}
  \end{subfigure}
  \begin{subfigure}{.45\textwidth}
    \includegraphics[width=7.8cm, height=6cm]{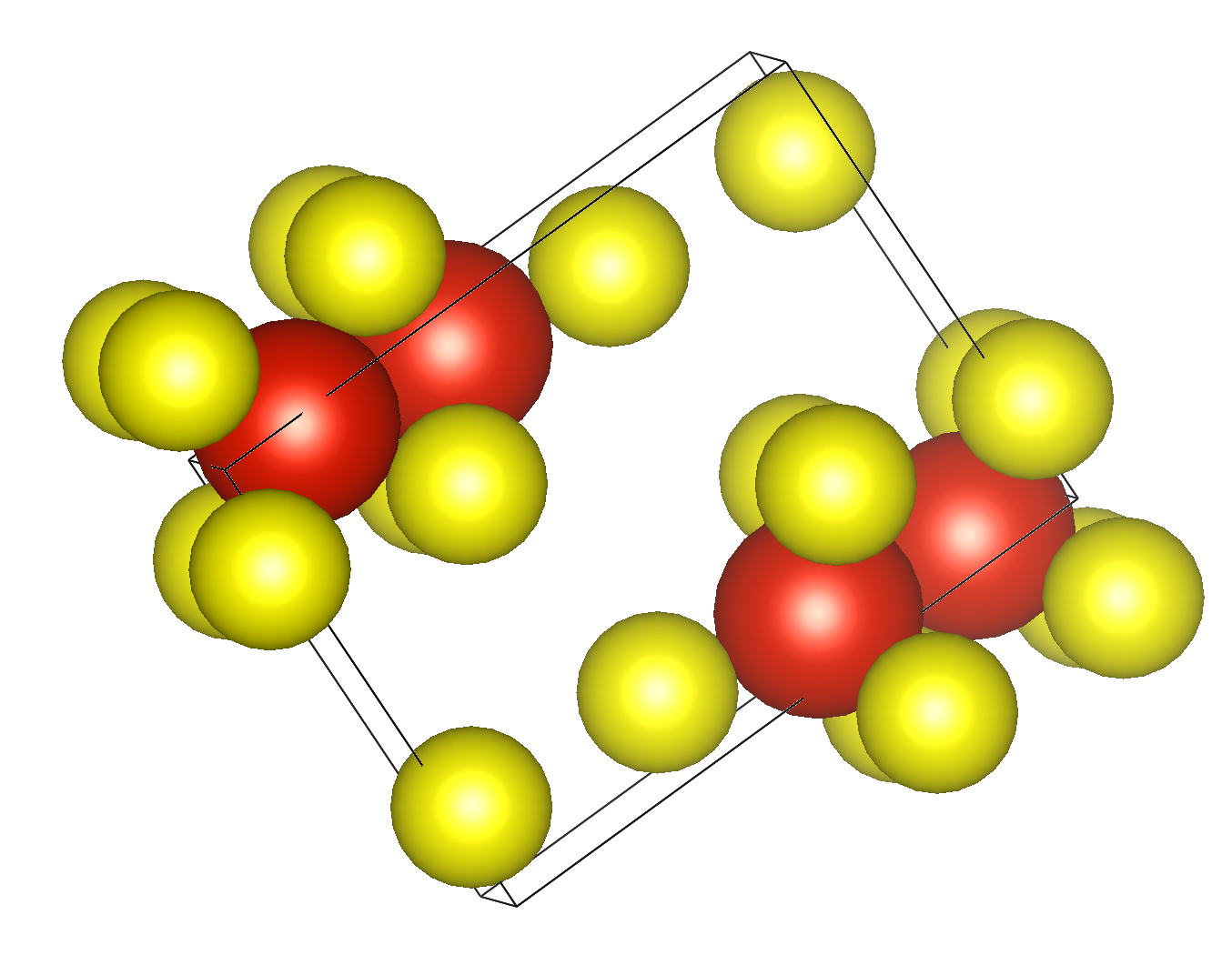}
    \caption{2D material V2S3 by element substitution with formation energy -0.907 eV}
    \vspace{3pt}
  \end{subfigure}
  \begin{subfigure}{.45\textwidth}
    \includegraphics[width=7.8cm, height=6cm]{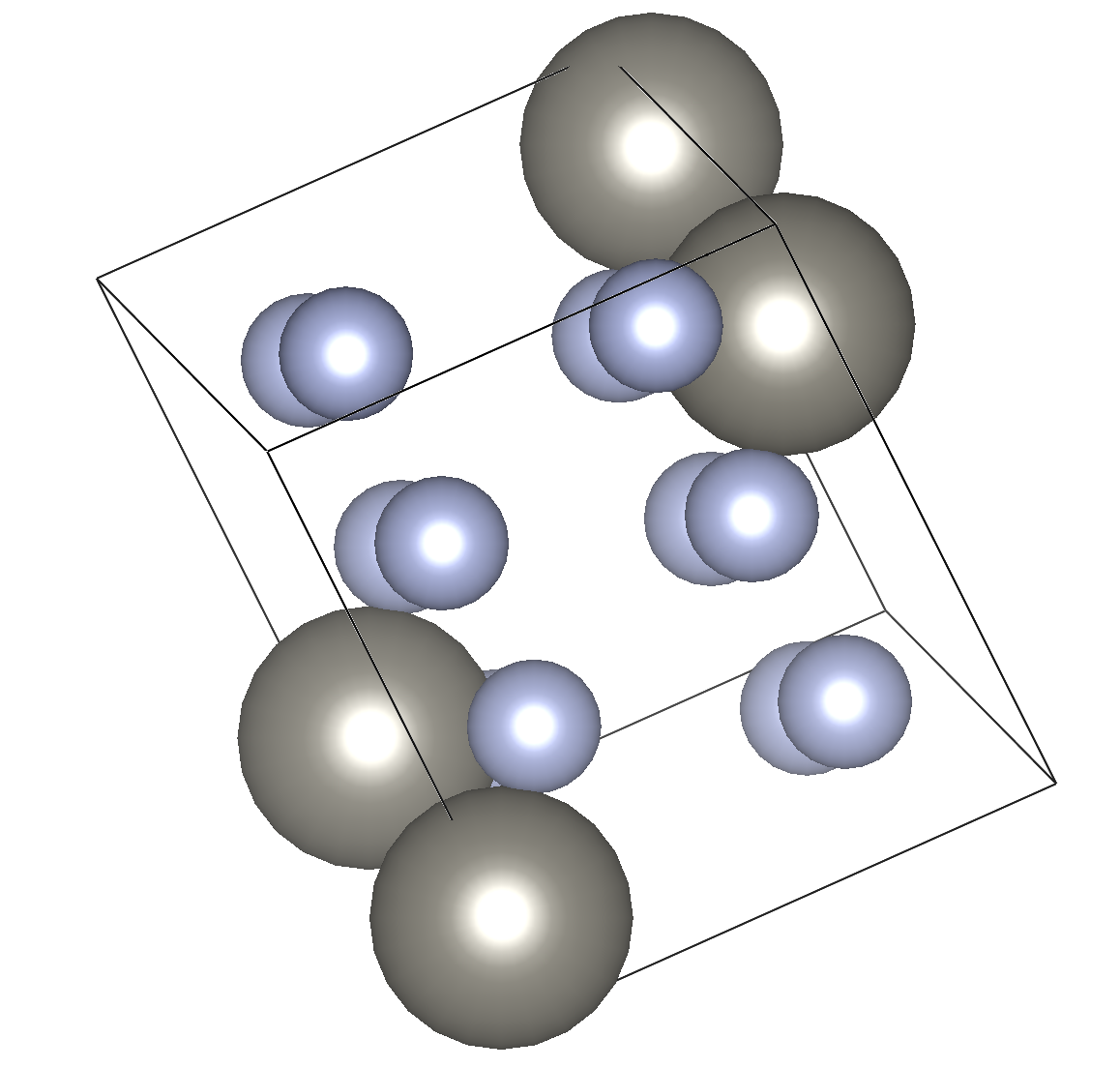}
    \caption{2D materials WF3 by element substitution with formation energy -2.255 eV}
    \vspace{3pt}
  \end{subfigure} 
  
  
  \caption{Selected structures of the discovered new 2D materials with DFT validation}
  \label{fig:structures_found}
\end{figure}

To verify the predicted 2D materials, we picked 1485 predicted material formulas with the highest probability scores and try to use the template based structure prediction method to find their structures. For 101 materials, we have found the templates of known layered materials or 2D materials in the 2DMatPedia database. We then predict the space group of our predicted 2D materials and choose the templates with the same or similar space groups. We then apply element substitution to get their structures. Some of the predicted 2D materials structures are shown in Figure\ref{fig:structures_found}. In total, 101 predicted materials with structures have been obtained. We then applied DFT relaxation to 17 of these structures and calculate their formation energies. In total we found 12 hypothetical materials with negative formation energy as shown in Table \ref{table:dft_formation_energy}. We found they are all binary materials with space group numbers ranging from 11 to 166. The structures of four of these materials are shown in Figure\ref{fig:structures_found}, all showing layered structures as expected.

To further verify whether these hypothetical materials can really exist, we applied DFT first principle calculation to compute the formation energies per atom for 12 2D-layered materials as shown in Table~\ref{table:dft_formation_energy}. The materials with high formation energies like TaF$_4$ (-2.8048 eV/atom) and WF$_3$ (-2.2548 eV/atom) implies that the proposed method in this research is able to discover 2D layered materials which are highly thermodynamically stable against the parent compounds of their elements.


\begin{table}[h!]
\centering
\caption{Hypothetical 2D materials with DFT verified formation energies in eV/atom}
\begin{tabular}{cccccc}
\hline
Formula & FormationEnergy & SpaceGroup & Formula & FormationEnergy & SpaceGroup \\ \hline
TaF4    & -2.8048         &   P$2$/c {[}15{]}          & YI2     & -1.2603         &   C2/m {[}12{]}         \\
WF3     & -2.2548         &   P$2_{1}$/c {[}14{]}         & YS3     & -1.1584         &  P$2_{1}$/c {[}14{]}          \\
YCl2    & -1.7472         &   C2/m {[}12{]}       & V2S3    & -0.9071         &   C2/m {[}12{]}          \\
MnCl3   & -1.4512         &   R$\overline{3}$m{[}166{]}           & MnBr3   & -0.6306         &   R3m{[}166{]}          \\
S2O5    & -1.4447         &   P$2_{1}$/m {[}11{]}        & AgCl2   & -0.5711         &     Pmmn {[}59{]}          \\
FB      & -1.3800         &   P$2_{1}$/c {[}14{]}           & SeSi    & -0.1462         &        P$\overline{3}$m1 {[}164{]}     \\ \hline
\end{tabular}
\label{table:dft_formation_energy}
\end{table}

\section{Conclusion}

We propose a generative inverse design approach for finding hypothetical new 2D materials. It includes a GAN based composition generation model for generating chemically valid materials formulas, a composition based random forest 2D materials classifier, a template based element substitution structure predictor, and DFT verification. Using this pipeline, we have generated 1485 hypothetical 2D materials with probability scores greater 95\%. We computationally verified 12 materials are 2D materials using DFT formation energy calculation and validation two materials' thermostability using DFT phonon calculation. These new hypothetical materials can be used to guide the screening of 2D materials for special functions using materials property prediction models. The experiments demonstrate the effectiveness of the proposed approach for discovering new 2D materials and can be used as a complement of the prototype based element substitution based generation approach.

\section{Contribution}
Conceptualization, J.H.; methodology, J.H., Y.S.; software, Y.S.; validation, Y.S., D.S. and J.H.;  investigation,J.H., Y.S., D.S., and Y.Z.; resources, J.H.; data curation, J.H., Y.S. and Y.Z.; writing--original draft preparation, J.H. and Y.S., E.S., and Y.Z.; writing--review and editing, J.H, Y.S.; visualization, Y.S.; supervision, J.H.;  funding acquisition, J.H.

\section{Acknowledgement}
Research reported in this work was supported in part by NSF under grant and 1940099 and 1905775. The views, perspective, and content do not necessarily represent the official views of the SC EPSCoR Program nor those of the NSF.

\bibliography{references}
\bibliographystyle{unsrt}

\end{document}